\documentclass[10pt,letterpaper]{article}
\usepackage[top=0.85in,left=2.75in,footskip=0.75in]{geometry}

\usepackage{amsmath,amssymb}

\usepackage{changepage}

\usepackage[utf8x]{inputenc}

\usepackage{textcomp,marvosym}

\usepackage{cite}

\usepackage{nameref,hyperref}

\usepackage[right]{lineno}

\usepackage{microtype}
\DisableLigatures[f]{encoding = *, family = * }

\usepackage[table]{xcolor}

\usepackage{array}

\newcolumntype{+}{!{\vrule width 2pt}}

\newlength\savedwidth

\raggedright
\usepackage[aboveskip=1pt,labelfont=bf,labelsep=period,justification=raggedright,singlelinecheck=off]{caption}

\makeatletter
\renewcommand{\@biblabel}[1]{\quad#1.}
\makeatother

\usepackage{lastpage,fancyhdr,graphicx}
\usepackage{epstopdf}
\fancyheadoffset[L]{2.25in}
\fancyfootoffset[L]{2.25in}
\begin{document}
\vspace*{0.2in}

\begin{flushleft}
{\Large
\textbf\newline{Incentivising cooperation by rewarding the weakest member} 
}
\newline
\\
Jory Schossau\textsuperscript{1},
Bamshad Shirmohammadi\textsuperscript{2}, and
Arend Hintze\textsuperscript{1,2,*}
\\
\bigskip
\textbf{1} BEACON Center for the Study of Evolution in Action, Michigan State University, East Lansing, Michigan, United States of America
\\
\textbf{2} Department of MicroData Analytics, Dalarna University, Falun, Dalarna, Sweden
\\
\bigskip

* ahz@du.se

\end{flushleft}
\section*{Abstract}
Autonomous agents that act with each other on behalf of humans are becoming more common in many social domains, such as customer service, transportation, and health care. In such social situations greedy strategies can reduce the positive outcome for all agents, such as leading to stop-and-go traffic on highways, or causing a denial of service on a communications channel. Instead, we desire autonomous decision-making for efficient performance while also considering equitability of the group to avoid these pitfalls. Unfortunately, in complex situations it is far easier to design machine learning objectives for selfish strategies than for equitable behaviors. Here we present a simple way to reward groups of agents in both evolution and reinforcement learning domains by the performance of their weakest member. We show how this yields ``fairer'' more equitable behavior, while also maximizing individual outcomes, and we show the relationship to biological selection mechanisms of group-level selection and inclusive fitness theory.



\section*{Introduction}
We are increasingly surrounding ourselves with artificial intelligence-controlled (AI) autonomous systems such as self-driving cars. These systems must be trained by some mechanism, which is usually reinforcement ~\cite{watkins1992q}, deep reinforcement learning ~\cite{sutton2018reinforcement}, or sometimes neuroevolution~\cite{stanley2019designing}.
Regardless of the optimization method, performance must be described by an objective function. With self-driving cars, the objective is to deliver the passengers safely from point A to point B, not damaging the car or others in the process, maybe saving energy, and to a large degree prioritizing arrival time -- the same objectives human drivers have. However, these objectives are selfish in that they ignore the success of others. A self-driving car optimized by these criteria would not be considerate in lane transition planning, intersections, or other potentially contentious situations in which humans have expressed consideration. This poses a simple question: How do we optimize autonomous decision-making AIs beyond their selfish objectives?

One way could be to include more criteria in the objective function. However, this leads to an explosion of criteria and does not necessarily address system efficiency. For example, each car yielding Right of Way to one waiting behind a stop sign is inefficient. Similarly, it might not be clear which extra objectives affect overall performance. Neither does it sound possible to quantitatively encode ``being considerate and optimal'' into an objective function.

Nature faces a similar problem during evolutionary optimization of decision-making. Evolution selects on short time rewards making it hard for the complex behavior of cooperation to evolve. Thus, different mechanisms have been proposed that could help cooperation to emerge ~\cite{nowak2006five}. The most obvious ones are kin selection ~\cite{queller1992general}, group selection ~\cite{smith1964group}, and inclusive fitness ~\cite{hamilton1964genetic}. Kin selection is a concept that is hard to transfer into the domain of reinforcement learning and thus we leave it out of the discussion here. In biology, group-level selection requires all members of the group to replicate together, let alone receive the same payoff. Inclusive fitness stipulates that the performance of one agent is dependent on another agent. Both these mechanisms find a way to award organisms both individually and from mutual support of other agents. Group-level and inclusive fitness selection improve cooperation, and emphasize the success of the group over the success of the individual. Thus, it should be possible to counter the emergence of selfish AI behavior by using group-level selection or inclusive fitness. 

Multiple examples of group-level selection exist, and have been identified as drivers of major transitions in evolution, such as the transition from single- to multi-cellular organisms~\cite{rose_germ_lines_2020} or social insects~\cite{eowilson_1968}. While driving cooperation, group-level selection often favors efficient division of labor. Examples include the specialization of soma and germline in multicellularity, or the specialization of queens and workers in social insects. Even in less strict situations where groups of organisms are working together in a synergistic fashion to receive higher benefits, the rewards are may not be distributed equally~\cite{owens1978feeding}. This division of labor from group-level selection creates a situation where members receive more than they would alone, but still encounter unequal rewards~\cite{bongard2000legion,potter2001heterogeneity}- typically described as a despotic distribution~\cite{andren1990despotic}.

Interestingly, the fitness function for optimization in a genetic algorithm can also be applied for optimization in reinforcement learning~\cite{Bloembergen_Tuyls_Hennes_Kaisers_2015}. The concept of group-level selection and inclusive fitness can thus be transferred to reinforcement learning. However, inheritance only plays a role in the context of a genetic algorithm wherein populations of agents compete and the fitter ones replicate proportionally more often. This is different from reinforcement learning wherein agents do not replicate. Instead of optimizing a single agent, a group of agents can be trained using an objective function rewarding group performance. This is either accomplished with one policy controlling the actions of all agents at the same time ~\cite{bucsoniu2010multi}, or by training independent agents that share information or experiences~\cite{tan1993multi}. Thus, inclusive fitness is more akin to using independent policies, while group-level selection is closer to the optimization of a single policy.

Group performance for foraging agents can be assessed by taking the summation of the individually collected rewards or by taking the maximum of those rewards. This typically leads to poor overall performance, diverse behavior, and a higher despotic index~\cite{balch1999reward}. Also, the distributed nature of learning poses problems to exploration and learning schedules~\cite{hu2003nash}. On the other hand, this heterogeneous outcome might be desired to solve other tasks~\cite{Panait_Luke_2005}. It has also been argued that global reward schemes do not scale to larger groups and that using individual reward schemes remedies this problem~\cite{wolpert2002optimal}.

This credit assignment difficulty can be summarized as an act of finding the balance between local individual reward that can cause counterproductive interference, and maximizing global group reward that can lead to self-sacrificial inefficiencies at the local scale. For this reason many adaptive methods under the name ``shaped reinforcement learning'' have been proposed~\cite{mataric1997reinforcement, Bongard_2000, Wang_Zhang_Kim_Gu_2020}, being only a few.
It seems that the literature suggests that a global reward scheme evaluating the maximal or joint effort of a group neither leads to optimal performance, nor does it flatten the despotic index. 

Regardless, here we show that assessing the performance of a group by its weakest member leads to optimal performance, while also resulting in a fair distribution of labor and reward -- a flat despotic index. We show this for both genetic algorithms and reinforcement learning. The task used here is a foraging task, and performance is optimized by either a genetic algorithm or by reinforcement learning of policies controlling groups of individuals or the entire group. Three different rewarding schemes are compared:

\begin{itemize}
    \item \bf{MEAN}: the resources are pooled and then fairly distributed among the four agents
    \item \bf{MINIMUM}: each agent gets the same score defined by the agent who collected the least food
    \item \bf{MAXIMUM}: each agent gets the same score defined by the agent who collected the most food (a control)
\end{itemize}

We will show that the MINIMUM reward scheme indeed leads to high performance while also satisfying a low despotic index.

\section*{Materials and Methods}
\subsection*{The Foraging Task}
The foraging task requires four agents to collect food at the same time, which requires no explicit cooperation. After the foraging period is over, the amount of food each agent has collected is evaluated and used to determine either fitness for the GA or rewards for RL. However, when training a policy using RL, the state and action space must be kept small for feasible runtimes, so we use a simplified version.


\subsubsection*{Environment using a Genetic Algorithm}
Four agents are placed in the corners of a rectangular room ($16 \times 16$ tiles). The room is filled with food that agents can collect. The agents can move forward onto food and automatically collect it, they can turn left or right, do nothing, and also utter a far-ranging beeping signal as a form of communication. Agents can further deposit already collected food in front of them, or if an agent is directly in front, the food gets handed to that agent. Agents receive inputs of how much food they collected, what is immediately in front of them, and the beeping signals of other agents.

\paragraph{Objective Function for the Genetic Algorithm}
The four agents to forage in the environment are either chosen randomly from the population (without replacement) and evaluated once (individual), or each agent in the population is cloned three times to create a group of four identical agents that are evaluated in the environment (clone). In either case, the performance of the group is determined using one of three statistical methods: MEAN, MAXIMUM, and MINIMUM (see above).

\paragraph{Markov Brain neural networks}
Agents are controlled by Markov Brains~\cite{hintze2017markov}, which are neural networks that replace the commonly used aggregation and threshold functions with probabilistic or deterministic logic gates as well as mathematical operators to compute the state of nodes. The exact connectivity and use of computational units is defined by a genome subject to point mutations, deletions, and gene duplications. These mutations allow the computational units to connect to seven sensor nodes, three actuators, and nine hidden nodes. Two of the five sensor nodes convey what lies in front of the agent, and three for the beeping of other agents. The two output nodes encode the movement of the agent (left, right, nothing, and moving forward/eating/giving food). The nine hidden nodes can store information in a recurrent fashion, and thus allow memory to be stored.

\paragraph{Parameters for the Genetic Algorithm}
At the beginning of an evolutionary experiment, 100 agents are randomly initialized. When modeling inclusive fitness, four agents are randomly selected without replacement each generation and evaluated as a group. This sampling is repeated four times per generation to ensure the proper evaluation of each agent. When group-level selection is modeled, each agent of the population is evaluated by first making three clonal copies. The four identical agents are then evaluated together. Afterwards, three of the agents are removed such that whatever the group did only influences the subsequent possible replication of one agent. Optimization is performed over $50,000$ generations using roulette wheel selection.

\subsubsection*{Environment using Q-learning}
The environment for Q-learning is smaller ($8 \times 8$) and agents can see the entire area. Empty tiles are encoded as $0$, food is $1$, other agents are $-1$, and the agent to be controlled is represented as $-0.5$. Thus, the state of the environment $s$ is a tuple of length 64. The continuous values are chosen, so that the world can be used for deep-Q learning as well. Agents in this environment move in four cardinal directions (north, south, east, west). When an agent tries to move to a tile occupied by another agent, then one piece of food is transferred from the moving agent to the stationary agent. Because all agents see each other, no further communication is necessary.

\paragraph{Rewards for Reinforcement Learning}
We distinguish between two possible cases of learning: decentralized and centralized policies. When decentralized, each agent has its own policy (Q matrix) which is reinforced independently of the other agents. The Q matrix defines the rewards for four action: moving up down, left, or right. When centralized, the four agents are controlled by a single policy in such a way that the action selected defines in which direction each agent goes. In that case, the Q matrix must define rewards for 256 ($4^4$) possible actions to choose from, as there are four agents each able to move independently in four directions. 

\paragraph{Q-learning}
Q-learning was performed by allowing agents to forage for $50$ steps in an environment saturated with food, defining one epoch. Actions were based on the predictions of their respective Q-matrix. An exploration probability $\epsilon$ starting with $1.0$ and a decay rate of $0.999$ was used (greedy-epsilon with decay). If an exploration step was taken, the actions were selected randomly. After each step, the rewards, actions, and state were recorded to allow later training by memory replay. Rewards were defined identically to the three objective functions (MEAN, MAXIMUM, and MINIMUM) as outlined above. Agents were trained after each epoch using memory replay, selecting $2,000$ memories randomly from up to $50,000$ recorded ones. Memories were recorded using a deque such that newest memories replaced oldest memories. New rewards were calculated using a variant of the Bellman equation (Equation \ref{equ:BellmanEquation}):

\begin{equation}
\begin{aligned}
    Q^{new}(s_{t},a_{t})& \leftarrow Q(s_{t},a_{t})
    +\alpha(r_{t}+ \gamma  \max_a{Q}(s_{t+1},a)
    - Q(s_{t},a_{t})),\label{equ:BellmanEquation}
\end{aligned}
\end{equation}

which uses a learning rate $\alpha$ of $0.8$ (Equation \ref{equ:BellmanEquationAlt}):
\begin{equation}
\begin{aligned}
    Q^{new}(s_{t},a_{t})& \leftarrow (1.0-\alpha) Q(s_{t},a_{t})
    +\alpha(r_{t}+ \gamma  \max_a{Q}(s_{t+1},a))\label{equ:BellmanEquationAlt}
\end{aligned}
\end{equation}

Optimization was performed for $10,000$ epochs, and on average policies stopped improving after $5,000$ generations. Each stochastic experiment was repeated 40 times.

\paragraph{Q-matrix pruning}
There are $4^{64}$ possible states for the environment of $8\times8$ tiles with each tile having a state of empty, food, other agent, or the agent itself. This implies a dense Q-matrix of $4^{64}\times256$ for the condition of a policy controlling all 4 agents. Obviously, we did not have that much available memory at the time of writing. For this reason we used a dictionary to create a sparse Q-matrix, similar to~\cite{nishio2018faster}. Policy consultations resulted in a dictionary query for a reward vector based on the hash of an integer representation of the state. For non-existent states a random reward vector was created. All states are tagged with the last accessed epoch as a timestamp. Training an agent's policy over $50,000$ steps would still overflow the memory of a conventional computer so the dictionary was always pruned of entries unused for $2,000$ epochs. Pruning less often did not change the results (data not shown). 

This pruning method assumes consistent starting locations, which reduced the space of possible states to explore to those that are reachable within the 50 steps of the exploration. There was also a large diversity of rewarding paths that made the rewarding state space exploration much less uniform.
One reason for this diversity was that agents were not rewarded for visiting empty already-visited tiles.
As a result, the policy quickly converged on more rewarding action sequences,
and the rarely-consulted states could be pruned from the policy.

In some Reinforcement Learning architectures the policies can be prohibitively large. Deep Q-learning (DQN) is one method to counter this problem. In DQN the policy is not encoded as a table mapping states to predicted rewards, but as a deep neural network estimating rewards from states. While DQN is a powerful method, it couples the learning efficiency of Q-learning to the learning efficiency of deep neural network backpropagation. for our purposes this secondary process might obfuscate results and thus it was not used here.

\section*{Results}
\subsection*{Optimization by Genetic Algorithm}
\begin{figure}[t]
\begin{center}
\includegraphics[width=0.9\textwidth]{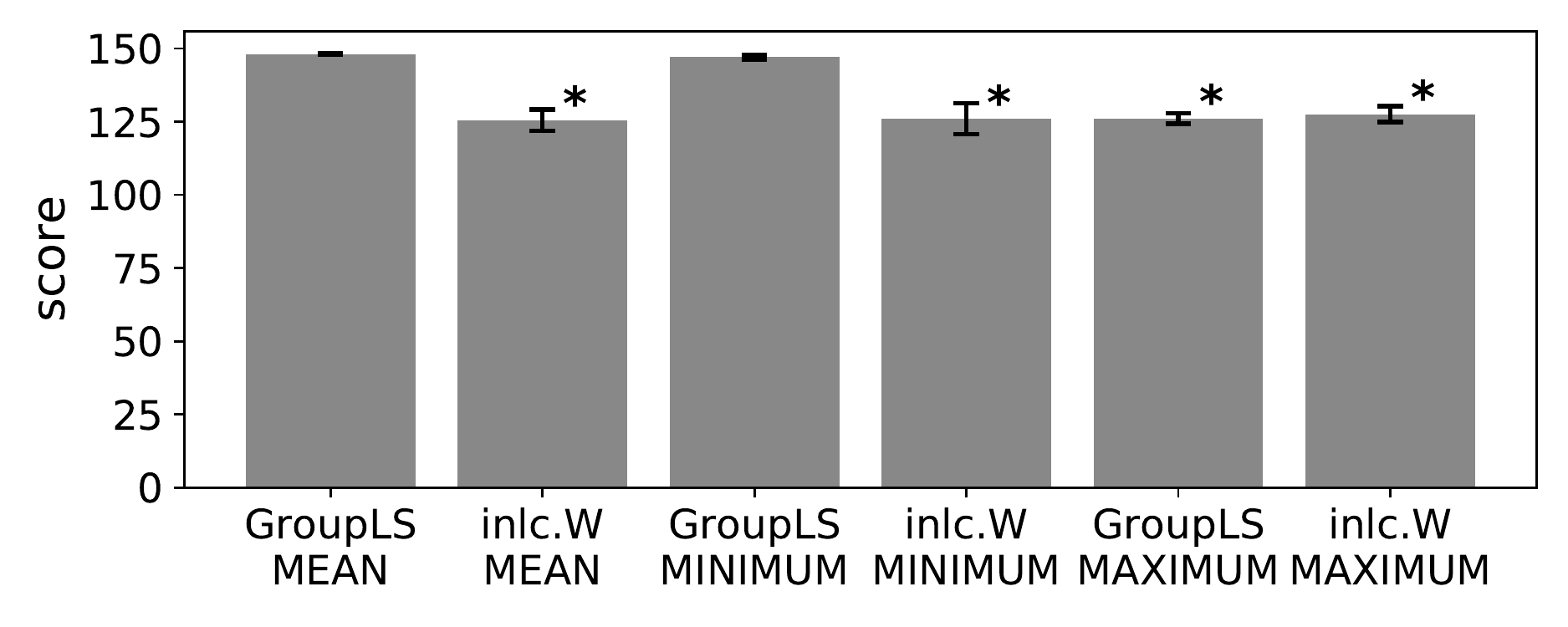}
\caption{
{\bf Average food collected by the entire group of agents optimized by a genetic algorithm, for different reward and selection regimes.} Group-level selection (GroupLS) uses clonal groups whereas Inclusive Fitness (Incl.W) uses individual selection. Error bars indicate 95\% confidence intervals over the $40$ replicates performed. $\**$ indicates that these results are statistically significantly different from Group-level selection using the MEAN and the MINIMUM reward scheme. P-value of the 2-sample Kolmogorov-Smirnov less than $\frac{0.01}{m}$, with $m$ as the number of samples when using Bonferroni correction to counter the multiple hypothesis testing error. Results from Group-level selection using MEAN and MINIMUM are not significantly different from each other.
}
\label{fig:GA_score}
\end{center}
\end{figure}

Agents performing in groups can either experience group-level selection when replicating as a group or experience inclusive fitness effects when replicating independently. To distinguish between both processes in our experiments we created groups for selection either from random population members (inclusive fitness effects) or clones of a single member (group-level selection).
We determined which group-selection method and reward scheme combination leads to the highest performance of a group for the reward schemes MEAN, MAXIMUM, and MINIMUM.
We find that all three reward schemes generally result in high performance, but that clonal groups using MEAN or MINIMUM reward schemes outperform the others (see Figure \ref{fig:GA_score}). This supports the hypothesis that the MINIMUM reward scheme in conjunction with group-level selection drives overall performance.

\begin{figure}[t]
\begin{center}
\includegraphics[width=0.9\textwidth]{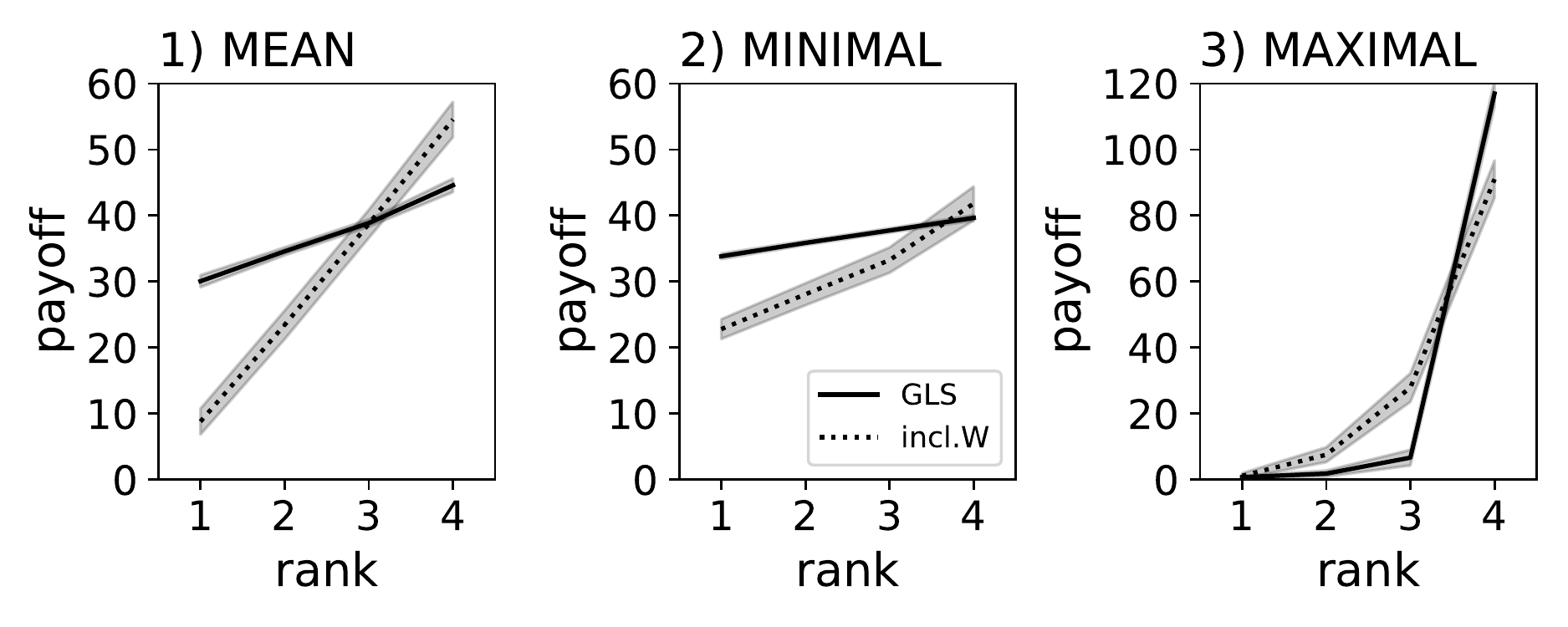}
\caption{
{\bf Individual food selected by agents trained using a genetic algorithm.} Solid lines indicate group-level selection using clonal groups (GLS), dotted lines indicate groups composed of genetically independent individuals experiencing only inclusive fitness effects (inlc.W). Individual results are sorted by rank (x-axis), which indicates amount of food collected relative to the other agents in the group. The gray shadows indicate the 95\% confidence intervals from the $40$ replicate experiments. Three different reward schemes were compared: MEAN, MINIMUM, and MAXIMUM - note the different y-scale for the MAXIMUM reward scheme.
}
\label{fig:GA_rank}
\end{center}
\end{figure}

We also investigated the effect of replication method and reward scheme on the despotic index.
As expected, the despotic index was highest when using the MAXIMUM reward scheme regardless of group-level selection or inclusive fitness (see Figure \ref{fig:GA_rank} 3).
The next steepest despotic index can be found when using the MEAN reward scheme, with group-level selection leading to a flatter hierarchy than inclusive fitness (see Figure \ref{fig:GA_rank} 1).
Finally, group-level selection with MINIMUM reward scheme resulted in a nearly equal resource distribution, indicating the most fair behaviors and outcomes for all agents in the group (see Figure \ref{fig:GA_rank} 2).
Using inclusive fitness and the MINIMUM reward scheme results in a flatter distribution of resources compared to MEAN and MAXIMUM, but is still steeper than group-level selection when using the MINIMUM reward scheme.

\subsection*{Optimization using Reinforcement Learning}
As discussed before, the concepts of group-level selection and inclusive fitness do not perfectly translate to reinforcement learning, since neither policies nor agents ``replicate''. However, using one policy to contemporaneously control all four agents can be equated with group-level selection --- called centralized control~\cite{Goldman_Zilberstein_2003}. In centralized control, rewards are used to reinforce all agent behaviors simultaneously. In contrast, inclusive fitness resembles using four independent policies --- called decentralized control. Rewards received by one agent only directly affect the policy of that agent, while the actions of other agents are only indirectly included through the reward scheme and lifetime interaction state changes.

\begin{figure}[t]
\begin{center}
\includegraphics[width=0.9\textwidth]{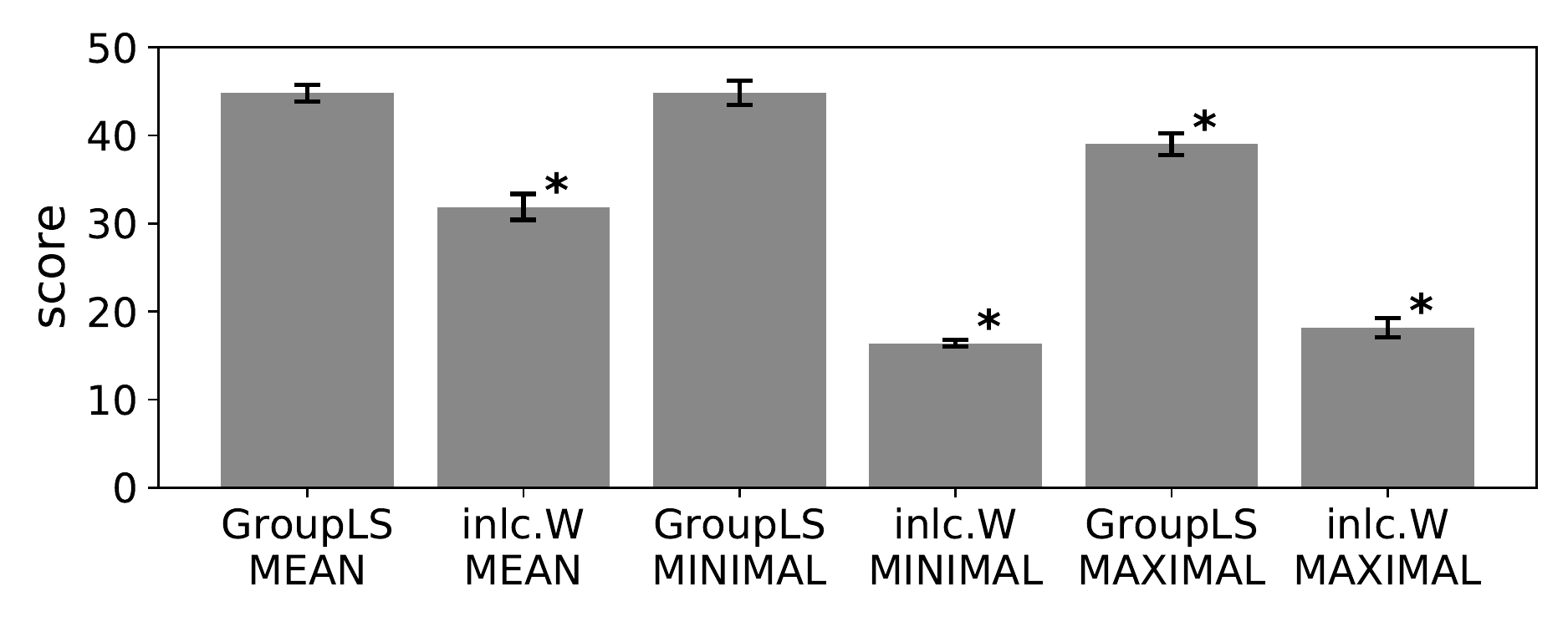}
\caption{
{\bf Average food collected by groups of agents optimized using Q-learning.} Group-level selection (GroupLS) indicate policies controlling all agents of the group at the same time, while groups of policies controlling only one agent at a time are analogous to inclusive fitness (inlc.W). Error bars indicate 95\% confidence intervals over the $40$ replicates performed. $\**$ indicate statistically significant difference from Group-level selection using the MEAN and the MINIMUM reward scheme. P-value of the 2-sample Kolmogorov-Smirnov less than $\frac{0.01}{m}$, with $m$ being the number of samples when using Bonferroni correction to counter the multiple hypothesis testing error. Group-level selection using MEAN or MINIMUM are not significantly different from each other.
}
\label{fig:Q_score}
\end{center}
\end{figure}

In all cases (MEAN, MINIMUM, and MAXIMUM) centralized control outperforms decentralized control, and 
MEAN and MINIMUM reward schemes yield better-performing agents than MAXIMUM when using centralized control.

\begin{figure}[t]
\begin{center}
\includegraphics[width=0.9\textwidth]{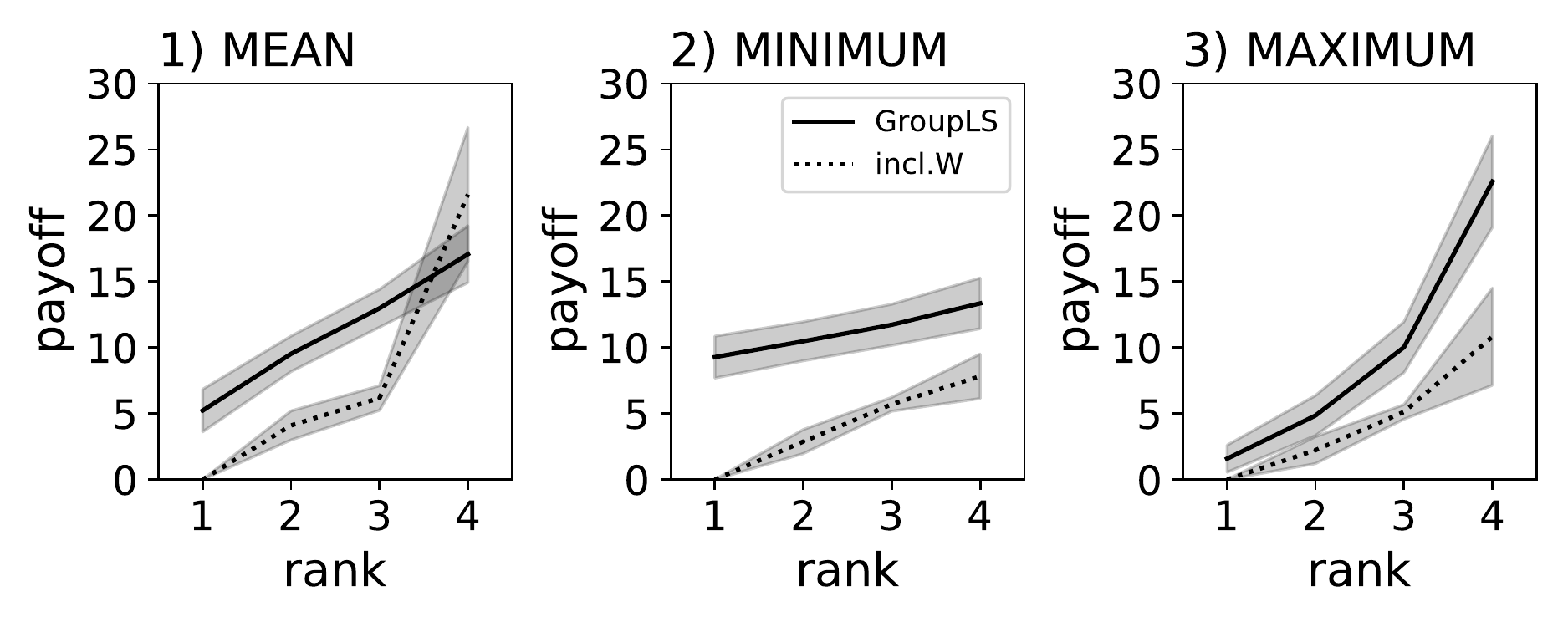}
\caption{
{\bf Individual food collected by agents controlled by Q-learning policies.} Solid lines show the centralized control group-level selection condition, where a single policy controls the actions of all group members(GroupLS). Dotted lines show groups composed of agents controlled by independent policies (inlc.W). Individual results are sorted by rank (x-axis) indicating relative success of food collection among the group. The gray shadows indicate the 95\% confidence intervals from the $40$ replicate experiments. Three different reward schemes were compared: MEAN, MINIMUM, and MAXIMUM.
}

\label{fig:Q_rank}
\end{center}
\end{figure}

Notably the despotic index is flattest when using the MINIMUM reward scheme and steepest when using the MAXIMUM reward scheme (see Figure \ref{fig:Q_rank}). In all cases, using decentralized (group-level) control policies resulted in flatter hierarchies and thus ``fairer'' agents.

\section*{Discussion}
Reinforcement learning and genetic algorithms are powerful tools that can effectively optimize agent behavior towards a predefined goal. However, designing the fitness function of criterion to achieve this goal is more an art than a science, especially for genetic algorithms. For instance, there is much research on how to quantify success for multiple objectives or difficult-to-define objectives with researchers not only engineering new mathematical algorithms, but drawing inspiration from grasshoppers, ant colonies, and immune systems, to name a few~\cite{mirjalili2018grasshopper,ning2019comparative,lin2018adaptive}. Being considerate to other members of a group while also maximizing personal reward is one of these cases. Agents who are incentivized to increase their personal rewards might do this on the expense of other agents. However, asking them to be considerate might in turn hamper their individual success. Balancing these two extremes is at the heart of this conundrum.

We found a simple reward scheme that satisfies these constraints and increases individual rewards while also promoting ``fairness`` among agents.
``Fairness`` --- the minimization of the despotic index --- is better achieved through rewarding a group by the performance of its weakest member, not by the average of all members or the success of its strongest member.
Group performance under this MINIMUM reward scheme is on par with rewarding the average, and is better than rewarding the best, at least for group-level selection. Similar maximization of a minimum objective in reinforcement learning has been shown to achieve decent performance~\cite{Tang_Song_Ou_Luo_Zhang_Wong_2020}.  However, the flattening of the despotic index resulting from this minimum objective was unknown. Here we study exactly this distribution effect of selection scheme at the interface of group-level selection and inclusive fitness.

Group-level selection and inclusive fitness are mechanisms known from biological evolution and thus translate easily to genetic algorithms.
Biological group-level selection is very similar to reinforcement learning of a single centralized control policy rewarded for group performance.
In a biological context, inclusive fitness describes the effect of other agents' behavior on an individual's fitness. Using different policies in reinforcement learning for different agents who receive independent rewards while still experiencing the effect of each other's actions resembles this biological mechanism. However, the exact degree to which these biological processes align with reinforcement methods remains open for debate.

The MINIMUM, MAXIMUM, and MEAN reward schemes were applied to Q-learning and not to deep Q-learning, which would replace the policy with a neural network to estimate future rewards based on the current state. However, our results should easily generalize to the deep Q-learning domain, as all principles tested here concern action-reward optimization, and not how well these rewards can be remembered.

The cooperative forging task used here allows for synergistic behaviors as well as antagonistic inefficient behaviors.
For the MAXIMUM reward scheme we found some agents collecting comparatively little food, thereby limiting overall group performance.
This is can be interpreted as a form of cooperation due to division of labor, where one agent is collecting resources while the others get out of the way. Alternatively, since agents can give resources to other agents, they may actively pool resources in one agent.
As such, the MAXIMUM reward scheme can not be directly compared to the MINIMUM reward scheme as they encourage two entirely different strategies.
However, this is exactly the result of this research: Using the MINIMUM reward scheme encourages an increase in every agent's performance in an equal fashion avoiding an unfair distribution.
This flattening of the despotic index is a consequence that pure individual MEAN or MAXIMUM reward schemes do not select for.
It would be interesting to test if the MINIMUM reward scheme is equally effective for economic and social structures to create fair but profitable resource distributions.

In practice, using the MINIMUM reward scheme might not even impose significant overhead. Imagine reinforcement training of a self-driving car in a virtual environment also simulating other cars. Using the minimum reward that any of the virtual cars obtains is a simple addition to such an simulation, because all car behaviors are modeled anyway.
Whether or not those other cars should be controlled by a centralized or decentralized policy, or if the MINIMUM reward scheme functions the same way in a mix of heterogeneous policies --- especially selfish ones --- remains an open question for future research.

\section*{Conclusion}

We proposed a simple way of minimum reward attribution that improves fairness in groups of agents trained by genetic algorithm or reinforcement learning. This biologically-inspired algorithm is interesting because it is easy to implement even in complex simulations and can work for either genetic algorithms or reinforcement learning, all while providing selection pressure or reinforcement signals for a fair distribution of agent population success.



\section*{Acknowledgments}
We thank Clifford Bohm for many insightful discussions. This work was supported in part by the BEACON Center for the Study of Evolution in Action at Michigan State University, and the National Science Foundation under Cooperative Agreement No. DBI-0939454. The computations were performed on resources provided by SNIC through Uppsala Multidisciplinary Center for Advanced Computational Science (UPPMAX) under Project SNIC 2020-15-48 and by the Institute for Cyber-Enabled Research at Michigan State University.

\nolinenumbers

%
%
%


\begin{thebibliography}{10}

\bibitem{watkins1992q}
Watkins CJ, Dayan P.
\newblock Q-learning.
\newblock Machine learning. 1992;8(3-4):279--292.

\bibitem{sutton2018reinforcement}
Sutton RS, Barto AG.
\newblock Reinforcement learning: An introduction.
\newblock MIT press; 2018.

\bibitem{stanley2019designing}
Stanley KO, Clune J, Lehman J, Miikkulainen R.
\newblock Designing neural networks through neuroevolution.
\newblock Nature Machine Intelligence. 2019;1(1):24--35.

\bibitem{nowak2006five}
Nowak MA.
\newblock Five rules for the evolution of cooperation.
\newblock science. 2006;314(5805):1560--1563.

\bibitem{queller1992general}
Queller DC.
\newblock A general model for kin selection.
\newblock Evolution. 1992;46(2):376--380.

\bibitem{smith1964group}
Smith JM.
\newblock Group selection and kin selection.
\newblock Nature. 1964;201(4924):1145--1147.

\bibitem{hamilton1964genetic}
Hamilton WD.
\newblock The genetic theory of social behavior. I and II.
\newblock Journal of theoretical biology. 1964;7:1--52.

\bibitem{rose_germ_lines_2020}
Rose CJ.
\newblock Germ lines and extended selection during the evolutionary transition
  to multicellularity.
\newblock Journal of Experimental Zoology Part B: Molecular and Developmental
  Evolution. 2020;doi:{https://doi.org/10.1002/jez.b.22985}.

\bibitem{eowilson_1968}
Wilson EO.
\newblock The Ergonomics of Caste in the Social Insects.
\newblock The American Naturalist. 1968;102(923):41--66.
\newblock doi:{10.1086/282522}.

\bibitem{owens1978feeding}
Owens MJ, Owens DD.
\newblock Feeding ecology and its influence on social organization in brown
  hyenas (Hyaena brunnea, Thunberg) of the central Kalahari Desert.
\newblock African Journal of Ecology. 1978;16(2):113--135.

\bibitem{bongard2000legion}
Bongard JC.
\newblock The legion system: A novel approach to evolving heterogeneity for
  collective problem solving.
\newblock In: European Conference on Genetic Programming. Springer; 2000. p.
  16--28.

\bibitem{potter2001heterogeneity}
Potter MA, Meeden LA, Schultz AC, et~al.
\newblock Heterogeneity in the coevolved behaviors of mobile robots: The
  emergence of specialists.
\newblock In: International joint conference on artificial intelligence.
  vol.~17. Citeseer; 2001. p. 1337--1343.

\bibitem{andren1990despotic}
Andren H.
\newblock Despotic distribution, unequal reproductive success, and population
  regulation in the jay Garrulus glandarius L.
\newblock Ecology. 1990;71(5):1796--1803.

\bibitem{Bloembergen_Tuyls_Hennes_Kaisers_2015}
Bloembergen D, Tuyls K, Hennes D, Kaisers M.
\newblock Evolutionary dynamics of multi-agent learning: A survey.
\newblock Journal of Artificial Intelligence Research. 2015;53:659--697.

\bibitem{bucsoniu2010multi}
Bu{\c{s}}oniu L, Babu{\v{s}}ka R, De~Schutter B.
\newblock Multi-agent reinforcement learning: An overview.
\newblock Innovations in multi-agent systems and applications-1. 2010; p.
  183--221.

\bibitem{tan1993multi}
Tan M.
\newblock Multi-agent reinforcement learning: Independent vs. cooperative
  agents.
\newblock In: Proceedings of the tenth international conference on machine
  learning; 1993. p. 330--337.

\bibitem{balch1999reward}
Balch T.
\newblock Reward and diversity in multirobot foraging.
\newblock Georgia Tech Library. 1999;.

\bibitem{hu2003nash}
Hu J, Wellman MP.
\newblock Nash Q-learning for general-sum stochastic games.
\newblock Journal of machine learning research. 2003;4(Nov):1039--1069.

\bibitem{Panait_Luke_2005}
Panait L, Luke S.
\newblock Cooperative Multi-Agent Learning: The State of the Art.
\newblock Autonomous Agents and Multi-Agent Systems. 2005;11(3):387--434.
\newblock doi:{10.1007/s10458-005-2631-2}.

\bibitem{wolpert2002optimal}
Wolpert DH, Tumer K.
\newblock Optimal payoff functions for members of collectives.
\newblock In: Modeling complexity in economic and social systems. World
  Scientific; 2002. p. 355--369.

\bibitem{mataric1997reinforcement}
Matari{\'c} MJ.
\newblock Reinforcement learning in the multi-robot domain.
\newblock In: Robot colonies. Springer; 1997. p. 73--83.

\bibitem{Bongard_2000}
Bongard JC.
\newblock The legion system: A novel approach to evolving heterogeneity for
  collective problem solving.
\newblock In: European Conference on Genetic Programming. Springer; 2000. p.
  16--28.

\bibitem{Wang_Zhang_Kim_Gu_2020}
Wang J, Zhang Y, Kim TK, Gu Y.
\newblock Shapley Q-value: A Local Reward Approach to Solve Global Reward
  Games.
\newblock In: Proceedings of the AAAI Conference on Artificial Intelligence.
  vol.~34; 2020. p. 7285--7292.

\bibitem{hintze2017markov}
Hintze A, Edlund JA, Olson RS, Knoester DB, Schossau J, Albantakis L, et~al.
\newblock Markov brains: A technical introduction.
\newblock arXiv preprint arXiv:170905601. 2017;.

\bibitem{nishio2018faster}
Nishio D, Yamane S.
\newblock Faster deep q-learning using neural episodic control.
\newblock In: 2018 IEEE 42nd Annual Computer Software and Applications
  Conference (COMPSAC). vol.~1. IEEE; 2018. p. 486--491.

\bibitem{Goldman_Zilberstein_2003}
Goldman CV, Zilberstein S.
\newblock Optimizing information exchange in cooperative multi-agent systems.
\newblock In: Proceedings of the second international joint conference on
  Autonomous agents and multiagent systems. AAMAS '03. Association for
  Computing Machinery; 2003. p. 137--144.
\newblock Available from: \url{https://doi.org/10.1145/860575.860598}.

\bibitem{mirjalili2018grasshopper}
Mirjalili SZ, Mirjalili S, Saremi S, Faris H, Aljarah I.
\newblock Grasshopper optimization algorithm for multi-objective optimization
  problems.
\newblock Applied Intelligence. 2018;48(4):805--820.

\bibitem{ning2019comparative}
Ning J, Zhang C, Sun P, Feng Y.
\newblock Comparative study of ant colony algorithms for multi-objective
  optimization.
\newblock Information. 2019;10(1):11.

\bibitem{lin2018adaptive}
Lin Q, Ma Y, Chen J, Zhu Q, Coello CAC, Wong KC, et~al.
\newblock An adaptive immune-inspired multi-objective algorithm with multiple
  differential evolution strategies.
\newblock Information Sciences. 2018;430:46--64.

\bibitem{Tang_Song_Ou_Luo_Zhang_Wong_2020}
Tang J, Song J, Ou J, Luo J, Zhang X, Wong KK.
\newblock Minimum throughput maximization for multi-UAV enabled WPCN: A deep
  reinforcement learning method.
\newblock IEEE Access. 2020;8:9124--9132.

\end{thebibliography}

\end{document}